# Radiative electronic bound states in the continuum from defects in semiconductors

Seong Yun Hong[1], Liang Z. Tan[2], Ki Hoon Lee[3], Youngho Kang[1,4], and Yeonghun Lee[1,5,*]

[1]Department of Intelligent Semiconductor Engineering, Incheon National University, Incheon 22012, Republic of Korea

[2]Molecular Foundry, Lawrence Berkeley National Laboratory, Berkeley, California 94720, USA

[3]Department of Physics, Incheon National University, Incheon 22012, Republic of Korea

[4]Department of Materials Science and Engineering, Incheon National University, Incheon 22012, Republic of Korea

[5]Department of Electronics Engineering, Incheon National University, Incheon 22012, Republic of Korea

[*]y.lee@inu.ac.kr

**ABSTRACT:** Continuum-buried defect states in semiconductors are generally expected to be optically inactive due to their strong coupling to continuum bands. Here, we show that such defects can instead host radiative electronic bound states in the continuum (BICs), using the silicon G-center as a prototypical example. Hybrid-functional first-principles calculations with a Hubbard $U$ correction reveal that a localized defect state, initially buried below the valence band maximum (VBM) in the ground state, undergoes exchange-driven energy-level reordering under optical excitation and shifts above the VBM. This exchange-induced transition suppresses nonradiative decay and enables robust radiative emission. By computing temperature-dependent nonradiative lifetimes and comparing them with experimental photoluminescence (PL) lifetimes, we quantitatively reproduce the observed temperature dependence of the emission. These results uncover a stabilization mechanism for continuum-embedded defect states and establish electronic BICs as a general paradigm for designing defect-based optical systems, including quantum emitters and qubits.

**KEYWORDS:** Density functional theory, Bound states in the continuum, Silicon G-center, Quantum defect, Quantum applications

Photonic bound states in the continuum (BICs) have attracted broad interest in photonic platforms, where symmetry or interference protects localized optical eigenmodes whose frequencies lie within the continuum, realizing leak-free, long-lived high-Q resonances embedded in continuum bands.[1–5] While photonic BICs are well established, realizing analogous states in electronic systems is considerably more challenging. In real materials, the lattice or defect structure cannot be freely designed, and unavoidable coupling to electronic and phononic degrees of freedom readily breaks the ideal decoupling from the continuum. Such interactions typically delocalize bound states and enable rapid nonradiative recombination. While a few theoretical or experimental studies[6–8] have reported signatures consistent with electronic BICs, establishing strict BIC behavior in electronic materials is often nontrivial. In particular, the realization of electronic BICs associated with point defects in semiconductors has not yet been established.

Here, we extend the BIC concept to semiconductors by identifying defect states that are embedded within the valence band (VB) or conduction band (CB) continuum, as illustrated in Figure 1A. Unlike conventional in-gap defects that are strongly localized and largely decoupled from host material electronic states, these continuum-embedded defects are partially localized and exhibit finite hybridization with the host bands. Their existence and stability raise fundamental questions regarding electronic localization and radiative processes in a strongly coupled continuum environment. This concept of electronic BICs can be clarified by contrasting it with resonant defects. In resonant defect systems, such as resonant impurity levels in thermoelectric semiconductors, impurity states can hybridize with the host band structure and contribute to transport.[9] By contrast, the electronic BIC discussed here remains localized and only weakly hybridized with the host continuum, despite being energetically embedded within it. This distinction places our work in the broader context of defect physics in semiconductors.

Defect-based quantum emitters in semiconductors have attracted significant attention due to their integrability with scalable platforms and compatibility with existing electronic and photonic technologies.[10–12] Recent experiments with the silicon G-center demonstrate that continuum-embedded defect states can remain stable and emit single photons.[13–24] The G-center emits single photons at 1280 nm in the telecom wavelength range, making it well suited for quantum communication applications. Structurally, it consists of two substitutional carbon atoms with an interstitial silicon atom between them ($C_s$-$Si_i$-$C_s$),[25] forming a defect with $C_{1h}$ symmetry that hosts one localized state within the band gap and another below the valence band maximum (VBM) (see Figure 1B).[11,22] This observation challenges the conventional assumption that defect states buried below the VBM are unsuitable for quantum emission due to their rapid nonradiative decay into the continuum.[26] Although this has motivated recent studies to consider continuum-embedded defect states as potential candidates for quantum emission,[27,28] the physical mechanism that enables such states to avoid fast recombination and remain optically active has not been clearly established.

In this letter, we show that the silicon G-center realizes an electronic BIC and that, despite being energetically buried below the VBM and only weakly hybridized with the continuum, its stable photon emission originates from an exchange-driven energy-level reordering of the defect states in the excited configuration. By combining hybrid-functional Heyd-Scuseria-Ernzerhof (HSE06) calculations with a Hubbard $U$ correction,[22] electron localization function (ELF) analysis,[29–32] and quantitative evaluation of nonradiative lifetimes, we demonstrate that the buried defect state remains spatially localized and protected against phonon-assisted decay. Our results show excellent agreement with experiment and provide a clear microscopic explanation for how continuum-embedded defect states can remain optically active. Beyond the specific case of the G-center, our findings establish a conceptual and computational framework for identifying similarly robust defect-based optical systems in other semiconductors.

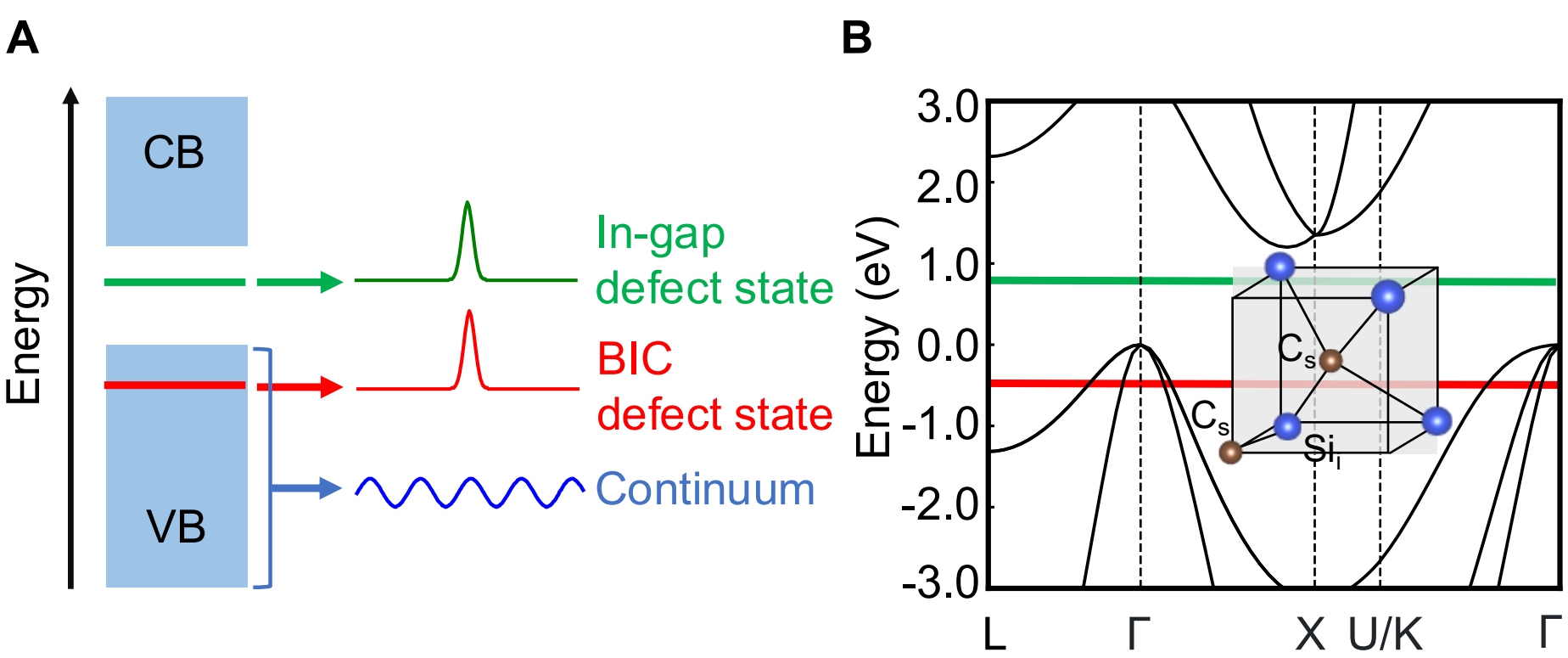

**Figure 1. Electronic BIC. (A)** Schematic illustration of an electronic BIC. **(B)** Hybrid-functional silicon band structure showing the BIC and regular in-gap defect states of the G-center, whose atomic structure shown in the inset. The defect levels are indicated by horizontal lines, extracted at the Γ point from the supercell calculation, where spin-up and spin-down

defect states are degenerate.

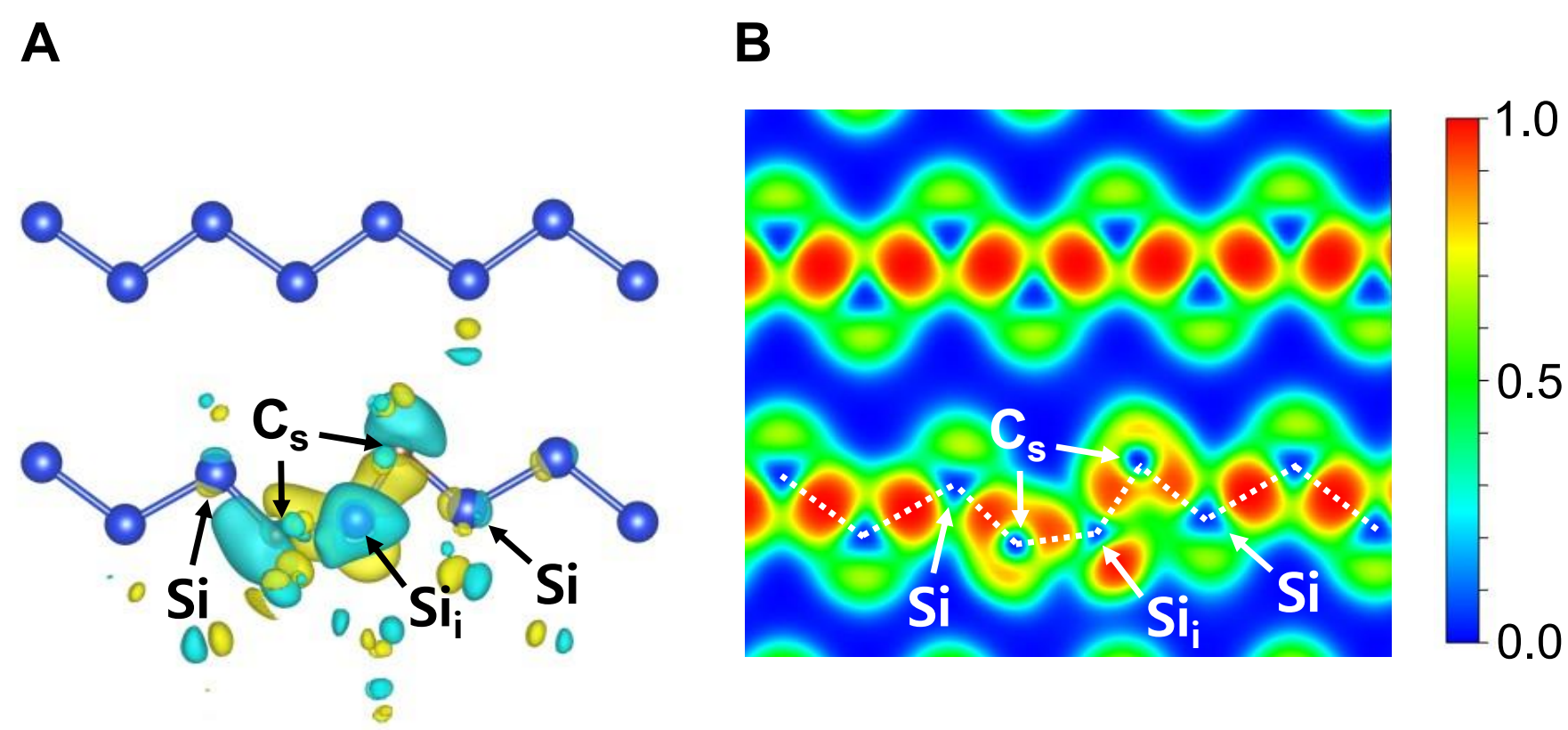

**Figure 2. Spatial localization of the BIC defect state.** (**A**) Wave function of the BIC defect state in the (110) plane, showing spatial localization. (**B**) ELF map. Compared with the Si-Si bonds, the Si-$C_s$ bonds exhibit reduced ELF values, indicating weakened covalent bonding associated with bond angle distortions.

**G-center as an electronic BIC**

The lower defect level of the silicon G-center in the ground state is buried below the VBM and remains spatially localized (see Figure 2A). This localization is not protected by symmetry. The silicon VBM at the $\Gamma$ point transforms as the $T_{2g}$ irreducible representation (neglecting spin),[33] which decomposes into $2A' + A''$ by incorporating the $C_{1h}$-symmetric G-center. As a result, hybridization with the $a'$ and $a''$ defect states of the $C_{1h}$ G-center is symmetry-allowed. Despite the symmetry-allowed mixing, the calculated defect orbital remains localized, indicating insufficient hybridization with the valence band continuum. The suppression of hybridization is first supported by the local bonding geometry along the successive Si-$C_s$-$Si_i$-$C_s$-Si bonding network. The Si-$C_s$-$Si_i$, $C_s$-$Si_i$-$C_s$, and $Si_i$-$C_s$-Si bond angles are found to be 124.4°, 130.1°, and 80.7°, respectively, deviating substantially from the ideal tetrahedral angle of 109.5° that maximizes $sp^3$ hybridization. Such angular distortions reduce orbital overlap and effective hopping between neighboring atoms, leading to a weakening of covalent hybridization along the defect pathway. As these bond angles progressively depart from 109.5°, the coupling between the defect orbital and the surrounding continuum states is suppressed, reinforcing its localization. This geometry-driven reduction of hybridization is conceptually analogous to the bandwidth-control mechanism in distorted rare-earth nickelates, $R\mathrm{NiO_3}$, where Ni-O-Ni bond angle distortions suppress hybridization, narrow the bandwidth $W$, enhance the effective correlation strength $U/W$, and ultimately drive a Mott transition.[34]

The ELF analysis provides direct evidence for this reduced hybridization. The ELF quantifies the conditional probability of finding two electrons with the same spin in proximity, where ELF = 1 indicates perfect localization (e.g., covalent bonding or lone pairs), while ELF ≈ 0.5 corresponds to delocalized electron gas behavior.[29–32] Along the first atomic row in Figure 2B, Si-Si bonds show ELF values near unity, indicating paired electrons localized between Si atoms and strong covalent character. In contrast, Si-$C_s$ and $C_s$-$Si_i$ bonds in the second atomic row exhibit reduced ELF values, where, like ionic bonding, electrons preferentially localized around the carbon atoms are less shared in interatomic regions, indicating a reduced covalent character. This continuous weakening of the covalency along the successive Si-$C_s$-$Si_i$-$C_s$-Si bonding network suppresses hybridization of the defect orbital with the continuum, allowing it to remain localized and form the electronic BIC. Furthermore, we evaluate the inverse participation ratio (IPR) to provide quantitative support for the localization of the continuum-embedded defect state, as shown in Figure 3 (see Figure S1 for the IPR analysis).

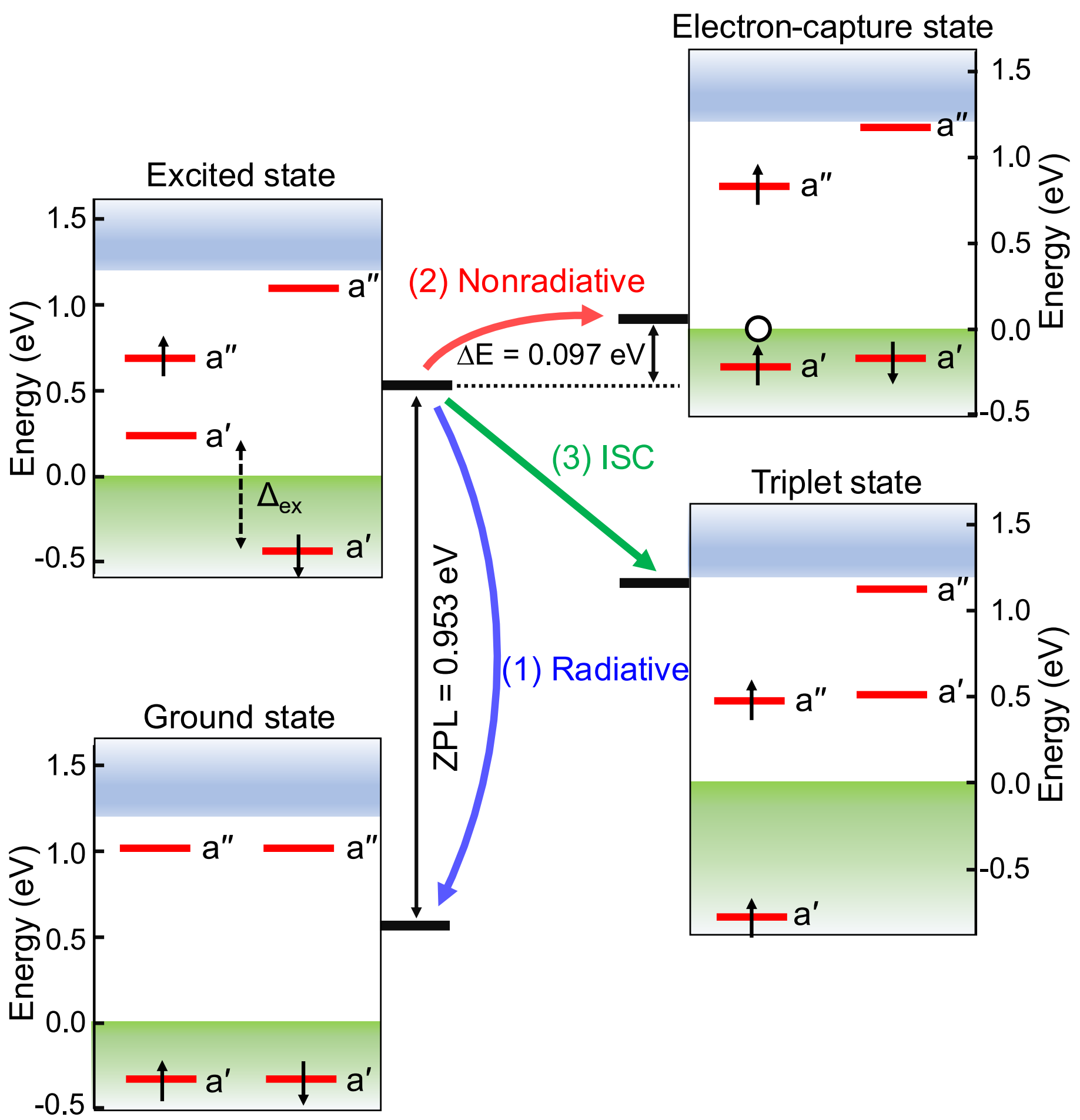

**Figure 3. Energy-level diagrams for the ground, excited, electron-capture and metastable triplet states.** The energy splitting $\boldsymbol{\Delta_{ex}}$ in the excited state denotes the exchange splitting, which lifts the $a'$ level above the VBM. In the excited state, three decay pathways are possible: (1) radiative recombination to the ground state with photon emission, (2) nonradiative recombination to the electron-capture state, where a valence band electron is transferred to the empty defect state via electron-phonon interaction, and (3) intersystem crossing (ISC) to the metastable triplet state.

**Exchange-driven energy-level reordering**

Spatial localization of a defect state alone is insufficient for photon emission if an efficient nonradiative decay channel into a continuum band exists. Figure 3 illustrates the defect energy levels of the silicon G-center under different electronic configurations. In the ground state, the lower defect level lies below the VBM, raising a fundamental question: how can such a continuum-embedded defect state become optically active once it is empty, given its energetic alignment with valence band states that would allow rapid nonradiative decay? The key mechanism is the exchange interaction that emerges in the excited state, where a single electron occupies a localized defect orbital. In this singly occupied state, the exchange interaction lowers the energy of the occupied orbital relative to the other, leading to a pronounced exchange splitting and stabilization of one spin state.[35] Consequently, the empty defect level that originally lies below the VBM in the ground state is shifted above the VBM in the excited state (see Figure 3), which can suppress phonon-assisted decay into the continuum.

To confirm the exchange-driven origin of this energy-level reordering, we performed additional calculations using the Perdew–Burke–Ernzerhof (PBE) functional.[36] In contrast to the hybrid functional HSE06+U (or HSE06), which incorporates nonlocal exchange, PBE fails to reproduce the reordering (see Figure S2 for more details), demonstrating that the excited state level shift originates from exchange. A similar exchange-driven energy-level reordering has also been reported for silicon W-centers, where it may be responsible for their optical activity.[37] This exchange-driven mechanism suppresses phonon-mediated electron transfer from the valence band to the empty defect state and provides the microscopic mechanism that enables stable single-photon emission from continuum-embedded defects.

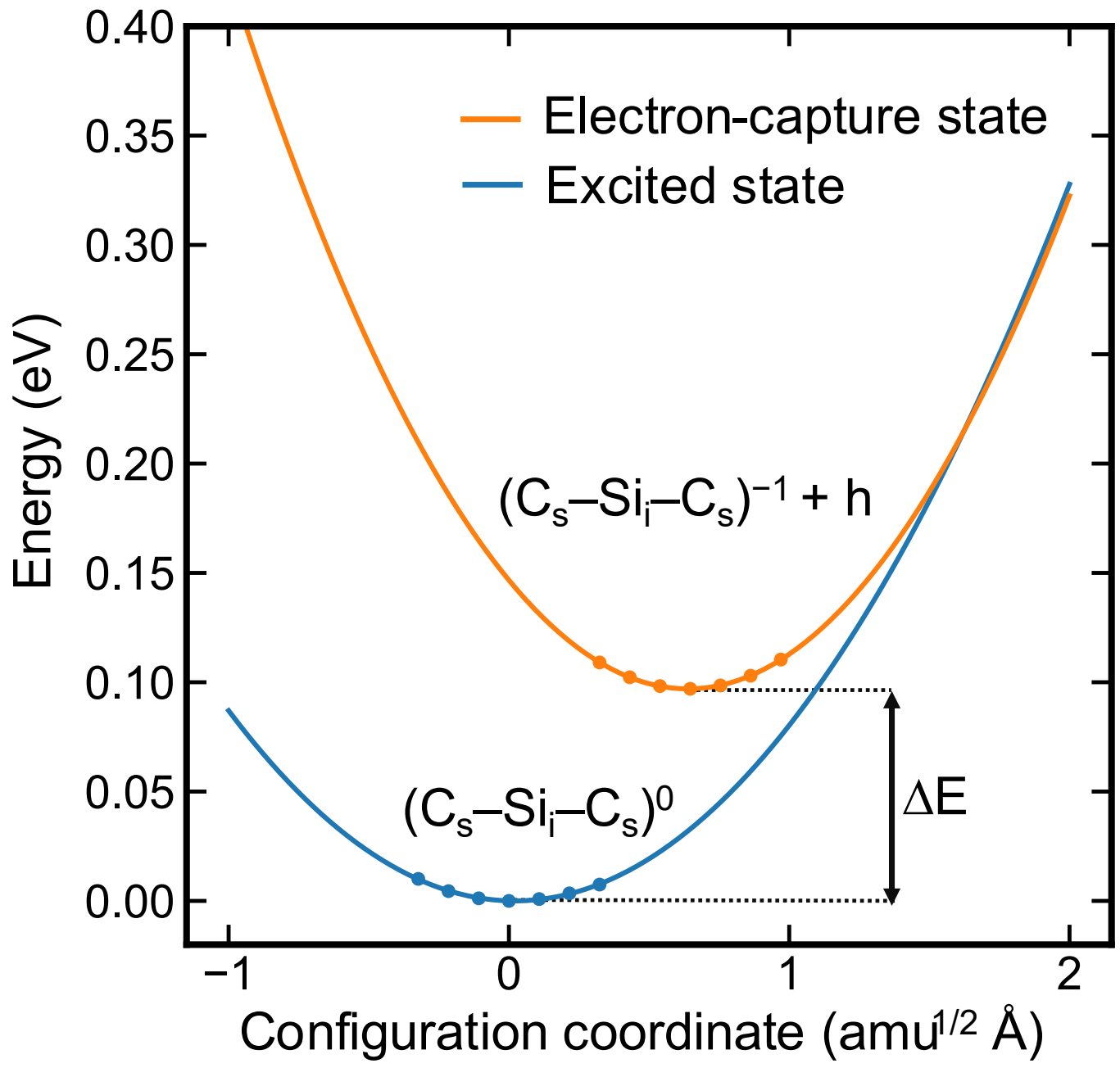

**Figure 4. Configuration coordinate diagrams of the excited and electron-capture states used to evaluate the nonradiative lifetime.** The hole capture coefficient $C_\mathrm{p}$ is obtained from the CCDs, and the hole emission rate, which corresponds to the electron capture rate, is calculated from the extracted $C_\mathrm{p}$.[38,39]

**Nonradiative lifetime**

To assess whether the G-center can remain optically active, we evaluate the nonradiative lifetime associated with phonon-assisted electron capture. The process involves two electronic configurations: the excited state, characterized by a singly occupied defect level slightly above the VBM, and the electron-capture state, in which this level is filled by an electron from the valence band (see Figure 3). Since this defect level lies close to the VBM, electrons in the valence band can be thermally promoted into the empty state via electron-phonon interactions. The nonradiative lifetime, therefore, is governed by the rate of this electron transfer process. This process competes with ISC into the metastable triplet state, which provides an additional relaxation pathway. To compute the nonradiative recombination rate, we adopt the hole emission picture,[39,40] in which the upward transfer of an electron from the valence band to the defect level is treated equivalently as the downward emission of a hole from the defect state to the valence band. Within this framework, the hole emission rate is given by

$$e_{\mathrm{p}} = C_{\mathrm{p}} N_{\mathrm{v}} \exp\left(-\frac{\Delta E}{k_{\mathrm{B}} T}\right),$$

where $C_{\mathrm{p}}$ is the hole capture coefficient, $N_{\mathrm{v}}$ is the temperature-dependent effective density of states at the VBM, and $\Delta E$ is the charge transition level between the excited and electron-capture states. The inverse of this rate defines the nonradiative lifetime.

A prerequisite for determining $\Delta E$ is an accurate description of the total energies of the relevant configurations. The excited state of the G-center is an open-shell singlet, which cannot be properly described within a single-determinant density functional theory (DFT). We, therefore, apply an open-shell singlet correction.[22,41,42] In addition, the singlet geometry is treated approximately within a single-determinant DFT using the excited state electronic configuration shown in Figure 3.[22,43] The validity of this approach is supported by the calculated zero-phonon line (ZPL) energy, defined as the total energy difference between the ground and excited states (see Table S1). We obtain a ZPL of 0.953 eV, in good agreement with the experimental value of 0.97 eV.[25,44] Using the corrected excited state energy, we evaluate the charge transition level $\Delta E$ of 0.097 eV (see Figure S3). The positive sign of $\Delta E$ has a clear physical meaning: it reflects that the excited state defect level has been shifted above the VBM due to the exchange effect (see Figure 3). This reordering, evident in both the Kohn-Sham energy levels and total energies, plays a central role in enabling stable photon emission from a continuum-embedded defect.

We obtain the hole capture coefficient, $C_{\mathrm{p}}$, by constructing configuration coordinate diagrams (CCDs) for the excited and electron-capture states (see Figure 4). Although the physical process considered here does not involve explicit hole capture, the CCDs provide a convenient and well-established means to extract the hole capture coefficient, $C_{\mathrm{p}}$, required to calculate the hole-emission rate. Supporting Information Figure S4 shows the calculation of electron-phonon coupling matrix elements for capture-coefficient evaluation. Using the computed values of $\Delta E$, $C_{\mathrm{p}}$, and $N_{\mathrm{v}}$, we obtain the nonradiative lifetime as a function of temperature, providing the first theoretical evaluation of the nonradiative lifetime of the silicon G-center.

For comparison with this nonradiative channel, we compute the radiative lifetime, $\tau_{\mathrm{r}} = 0.68\ \mu\mathrm{s}$, which is consistent with a previous first-principles study.[20] This comparison enables a direct assessment of the viability of photon emission from the G-center. The radiative lifetime of the silicon G-center is longer than that of other prominent in-gap defect-based quantum emitters, such as the diamond NV center (13 ns).[45] The lifetime difference arises from the lower ZPL energy (0.953 eV for the G-center compared to 1.945 eV for the NV center) and the reduced transition dipole moment (1.74 Debye for the G-center compared to 5.2 Debye for the NV center). Importantly, the smaller yet comparable dipole moment indicates that optical-transition-related defect orbitals remain spatially localized, rather than becoming delocalized into the host valence band continuum.

Experimentally, the photoluminescence (PL) lifetime of the G-center is measured to be 5.9 ns, and the PL intensity remains nearly temperature-independent between 10 K and 50 K.[13] This behavior reflects that the PL decay in the silicon G-center is predominantly governed by the ISC relaxation pathway involving the metastable triplet state (see Figure 3).[17,20,24] Previous work suggested that this ISC is allowed by spin-orbit coupling between the excited singlet and triplet states.[22] This spin-orbit-mediated ISC provides the microscopic pathway from the optically active excited singlet state to the metastable triplet state, which is treated as the dominant additional relaxation channel through the experimentally extracted ISC lifetime used in Figure 5. The characteristic time constant of this triplet-mediated decay pathway is extracted as the bunching time, $\tau_b$, from the second-order autocorrelation function, with reported values ranging from 6 ns to 15 ns.[17,24,46] In the following, we identify $\tau_b$ with the ISC lifetime, which naturally explains the nearly constant PL intensity observed at low temperatures. The dominance of the metastable triplet decay pathway is further reflected in the low experimental quantum efficiency (QE) of the silicon G-center at cryogenic temperatures, reported to be below 1%.[20] This is consistent with our calculated QE values of 0.85–2.1%, derived from the radiative lifetime and the ISC lifetime.

Figure 5 compares the temperature-dependent nonradiative lifetime with the radiative lifetime and the ISC lifetime. A direct comparison with the radiative lifetime yields a crossover near 73 K. However, because PL decay is governed primarily by the triplet-mediated pathway around the temperature, the physically relevant comparison is between the nonradiative lifetime and the ISC lifetime. This analysis reveals a crossover near 100 K, above which nonradiative recombination dominates. Below those temperatures, radiative recombination becomes favorable, indicating that photon emission from the silicon G-center is viable at cryogenic temperatures. Experimentally, the PL intensity decreases rapidly between 50 K and 110 K.[13,47] The calculated crossover temperature is in good agreement with experimental observations.[13,47]

We note that this comparison is subject to the approximations used in the excited state treatment. The excited state of the G-center is an open-shell singlet, which cannot be properly described within a single-determinant DFT because this state is not represented by a single Slater determinant but by a linear combination of determinants. We therefore evaluate its energy using the open-shell singlet correction, $E(S) = 2E(S/T) - E(T)$, within the ΔSCF framework.[12,41,42] Where $E(S/T)$, the mixed-spin ΔSCF state, is the excited state energy obtained directly from the practical single-determinant ΔSCF calculation for the open-shell excitation. Since the mixed-spin ΔSCF state contains equal singlet and triplet $m_S = 0$ character, this correction removes the triplet contribution and recovers the pure excited state singlet energy.[42] In addition, the singlet geometry is treated approximately within a single-determinant DFT using the excited state electronic configuration shown in Figure 3, assuming that the mixed-spin and excited state singlet states have similar equilibrium geometries. Although at these two points, uncertainty error may be introduced into the excited state treatment using DFT framework, our calculations are in good agreement with the experiment.

These results confirm that the exchange-driven energy level reordering plays a critical role in enabling radiative behavior from a defect state buried in the valence band continuum. The good agreement between the simulated and experimentally observed crossover temperatures demonstrates that incorporating the exchange-driven mechanism is essential to accurately describe the excited state dynamics of the silicon G-center. While the exchange effect suppresses phonon-assisted decay by energetically decoupling the excited defect level from the valence band, efficient photon emission is determined by competition among radiative recombination, nonradiative recombination, and other decay pathways, such as ISC. The interplay of these relaxation pathways explain how the continuum-embedded defect of the G-center can exhibit stable emission at cryogenic temperatures.

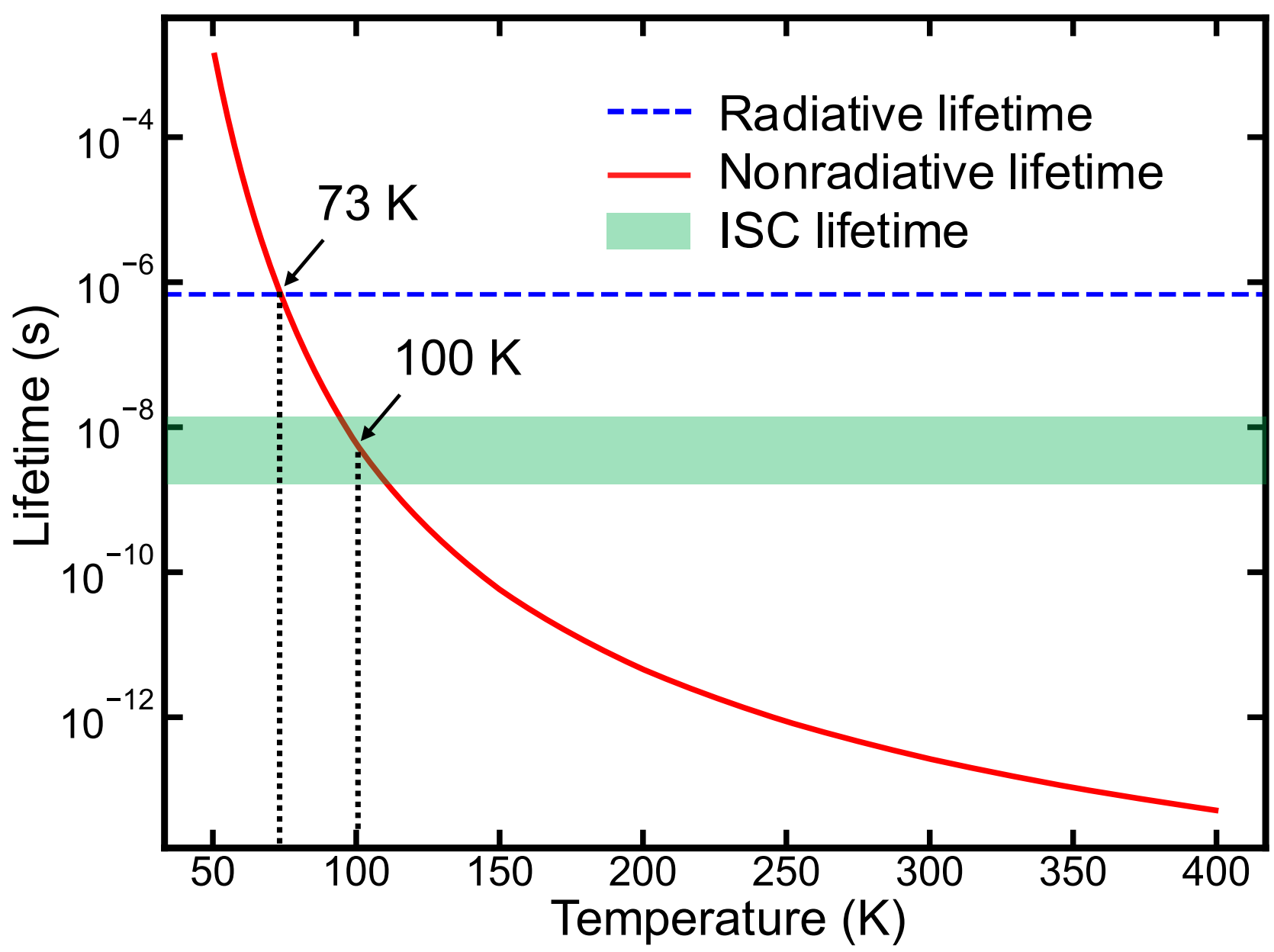

**Figure 5. Comparison of the radiative lifetime and ISC lifetime with temperature-dependent nonradiative lifetimes.** The shaded region indicates the experimentally reported ISC lifetimes ranging from 6 ns to 15 ns.[17,24,46] A crossover between the nonradiative and radiative lifetimes is observed at 73 K, while a crossover between the nonradiative and ISC lifetimes occurs around 100 K.

The silicon G-center exemplifies how a defect state embedded in the host continuum, i.e., an electronic BIC that remains localized because of weak hybridization rather than a strictly symmetry-protected BIC mechanism, can become optically active. The G-center can emit photons despite the defect level being buried below the VBM in the ground state because the exchange-driven energy-level reordering in the excited state lifts the empty defect level above the VBM and suppresses phonon-assisted electron transfer from the valence band. Despite overlapping with the valence band continuum in the ground state, the defect orbital remains localized and only weakly hybridized, as supported by bonding geometry and ELF analysis. Configuration coordinate diagrams yield temperature-dependent nonradiative lifetimes that cross the radiative lifetime and the ISC lifetime, indicating efficient emission at cryogenic temperatures, in agreement with experiments. Beyond establishing the G-center as a viable single-photon emitter, this work identifies a general route to suppress nonradiative decay in BIC defect states and opens new opportunities to discover and engineer robust defect-based optical systems, even in relatively small-band-gap semiconductors such as silicon.

**First-principles calculations**

First-principles calculations were performed using the Vienna Ab initio Simulation Package (VASP)[48,49] based on DFT.[50,51] The HSE06 hybrid functional,[52,53] combined with a Hubbard $U$ correction in the Dudarev approach[54] on the p orbital of the frustrated silicon self-interstitial atom ($U = 7.3$ eV), was employed.[22,54] The $U$ parameter was determined by aligning a defect level with a result from many-body perturbation theory with the GW approximation,[22] where the $U$ correction is relatively small and does not introduce a qualitative change in the defect physics, but improves agreement with experimental quantities such as the ZPL energy and charge-transition level (see Table S1 for functional dependence of related energy differences). A plane-wave cutoff of 400 eV was used. A 3×3×3 silicon supercell containing 216 atoms was used for the pristine structure, and the Brillouin zone was sampled using a single $\Gamma$-point. The defective structure was relaxed until the maximum atomic force was smaller than 0.01 eV/Å with fixed cell shape and volume of the pristine supercell. Excited states were calculated using the ΔSCF method.[12] The target orbitals for occupation constraints should be carefully selected because energy-level reordering occurs after excitation (see Figure S5 for the identification of defect orbitals). Charged-defect energies were corrected using the Freysoldt–Neugebauer–Van de Walle (FNV) scheme.[55] We used the Nonrad code to obtain hole capture coefficients for nonradiative lifetime calculations.[38,56]

**Corresponding Authors**

**Yeonghun Lee -** Department of Intelligent Semiconductor Engineering, Incheon National University, 119 Academy-ro, Yeonsu-gu, Incheon 22012, Republic of Korea; Department of Electronics Engineering, Incheon National University, Incheon National University, 119 Academy-ro, Yeonsu-gu, Incheon 22012, Republic of Korea; Research Institute for Engineering and Technology, Incheon National University, 119 Academy-ro, Yeonsu-gu, Incheon 22012, Republic of Korea; https://orcid.org/0000-0002-6058-1316; Email: y.lee@inu.ac.kr

**Authors**

**Seong Yun Hong -** Department of Intelligent Semiconductor Engineering, Incheon National University, 119 Academy-ro, Yeonsu-gu, Incheon 22012, Republic of Korea; https://orcid.org/0009-0006-8460-5518

**Liang Z. Tan -** Molecular Foundry, Lawrence Berkeley National Laboratory, Berkeley, California 94720, USA; https://orcid.org/0000-0003-4724-6369

**Ki Hoon Lee -** Department of Physics, Incheon National University, 119 Academy-ro, Yeonsu-gu, Incheon 22012, Republic of Korea;

**Youngho Kang -** Department of Materials Science and Engineering, Incheon National University, 119 Academy-ro, Yeonsu-gu, Incheon 22012, Republic of Korea; https://orcid.org/0000-0003-4532-0027

**Author contributions**

Y.L. conceived the idea and supervised the project. S.Y.H. and Y.L wrote the manuscript with input from all authors. S.Y.H. performed the first-principles calculations. All authors contributed to the development of the ideas, discussed the results, and commented on the manuscript.

**Notes**

The authors declare no competing interests. Supporting Information for this paper is available.

**Acknowledgements**

The authors thank Prof. Jeongwoo Kim for fruitful discussions and for fostering intellectual interactions. This work was supported by K-CHIPS(Korea Collaborative & High-tech Initiative for Prospective Semiconductor Research) (2410012153, RS-2025-02310666, 25073-15FC) funded by the Ministry of Trade, Industry & Energy (MOTIE, Korea) and by Basic Science Research Program through the National Research Foundation of Korea(NRF) funded by the Ministry of Education(RS-2025-25437019). Work at the Molecular Foundry was supported by the Office of Science, Office of Basic Energy Sciences, of the U.S. Department of Energy under Contract No. DE-AC02-05CH11231.

Supporting Information

# Radiative electronic bound states in the continuum from defects in semiconductors

Seong Yun Hong[1], Liang Z. Tan[2], Ki Hoon Lee[3], Youngho Kang[1,4], and Yeonghun Lee[1,5,*]

[1]Department of Intelligent Semiconductor Engineering, Incheon National University, Incheon 22012, Republic of Korea

[2]Molecular Foundry, Lawrence Berkeley National Laboratory, Berkeley, California 94720, USA

[3]Department of Physics, Incheon National University, Incheon 22012, Republic of Korea

[4]Department of Materials Science and Engineering, Incheon National University, Incheon 22012, Republic of Korea

[5]Department of Electronics Engineering, Incheon National University, Incheon 22012, Republic of Korea

[*]y.lee@inu.ac.kr

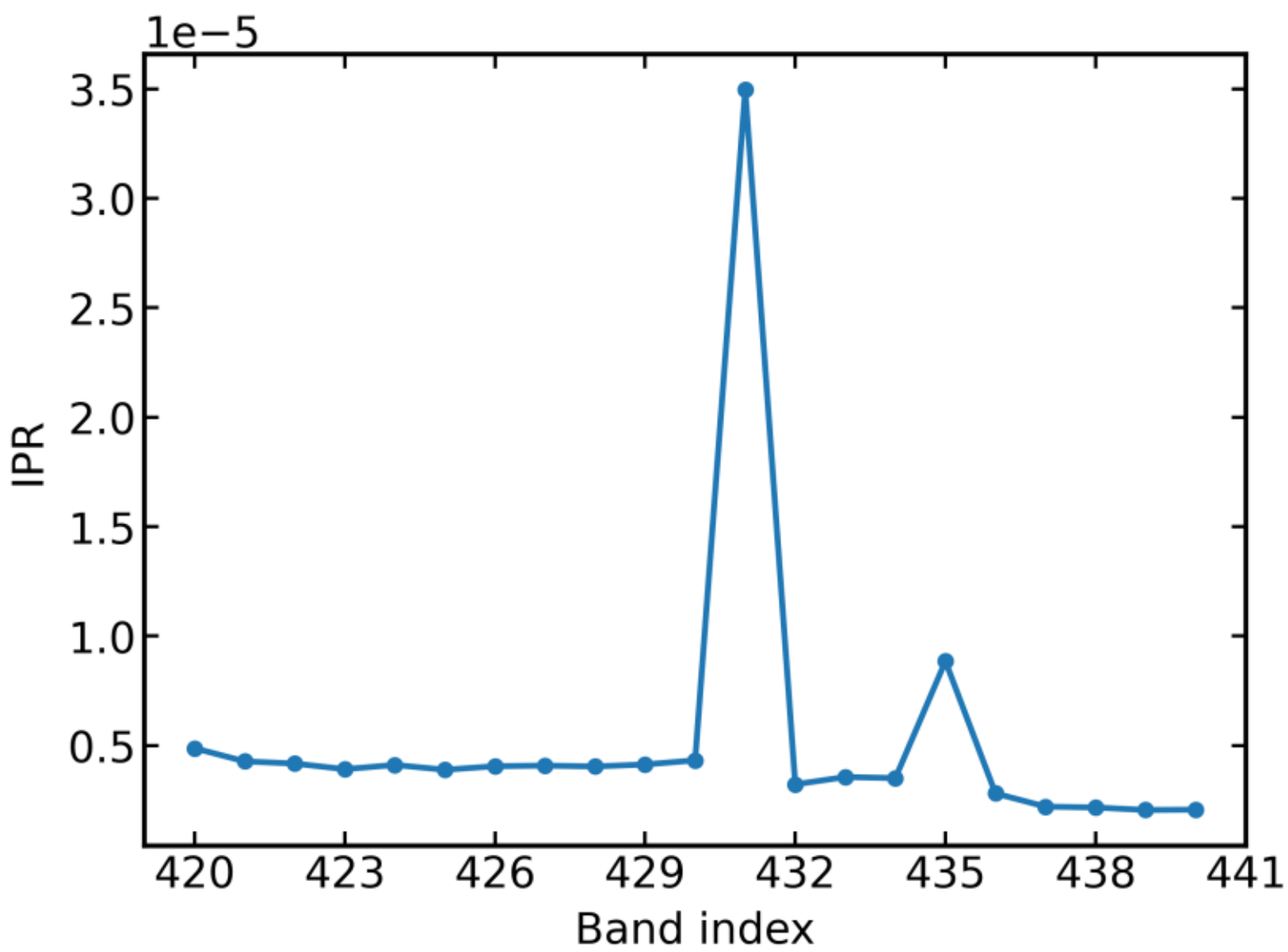

**Figure S1 Inverse participation ratio (IPR) of the ground state wavefunctions**. Distinct peaks appear at band indices 431 and 435, corresponding to the $a'$ and $a''$ defect states, indicating their localized character relative to the neighboring bands. The IPR is calculated from the Kohn-Sham pseudo-wavefunction on the real-space grid.

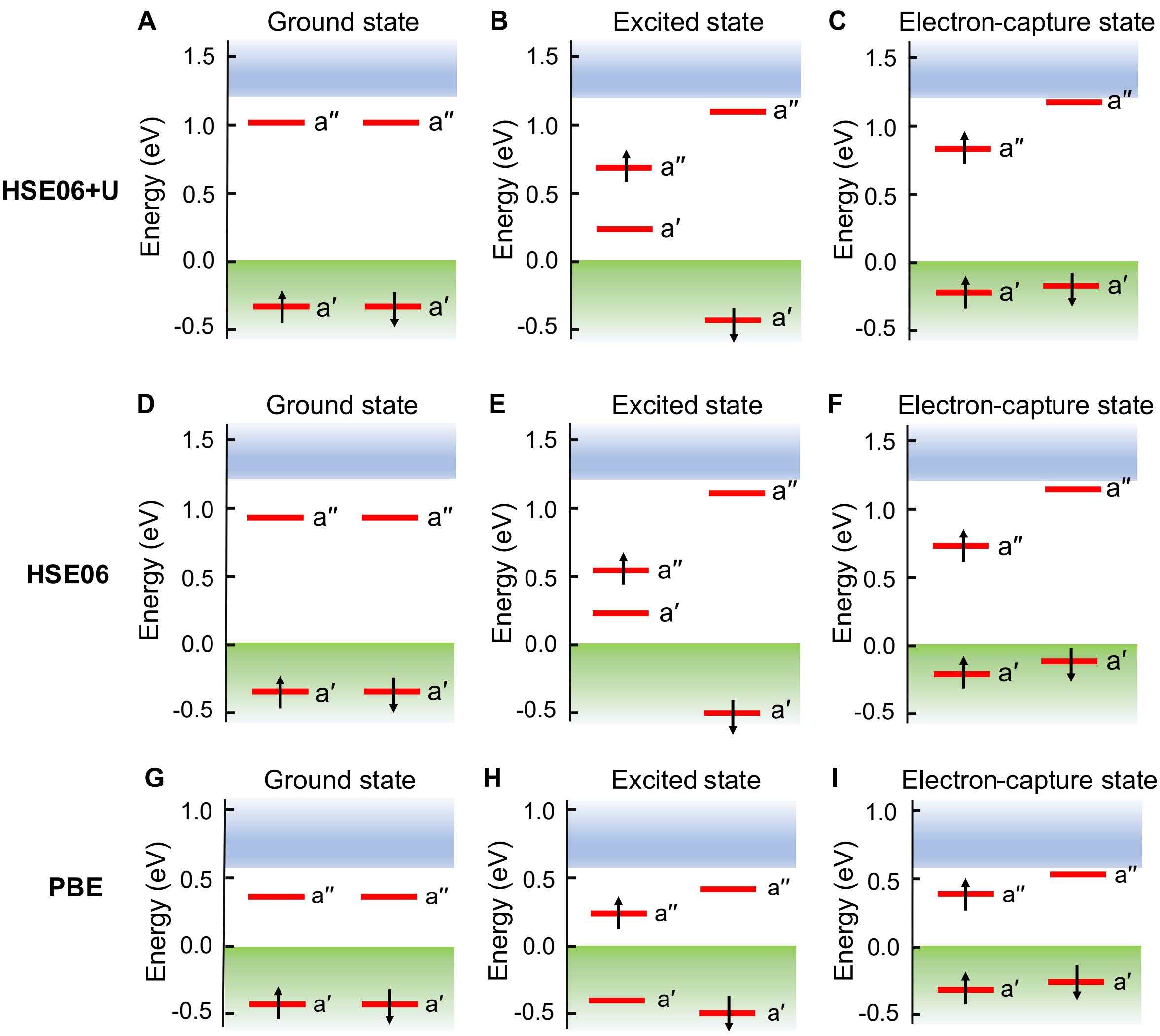

**Figure S2 Electronic structures in HSE06+U, HSE06, and PBE.** Electronic structures of the three different states using the HSE06+U (**A**, **B**, and **C**), HSE06 (**D**, **E**, and **F**), and PBE (**G**, **H**, and **I**). The exchange-driven energy-level reordering (EER) observed with HSE06+U and HSE06 is absent in the PBE results, supporting that the excited state defect-level shift originates from exchange effects.

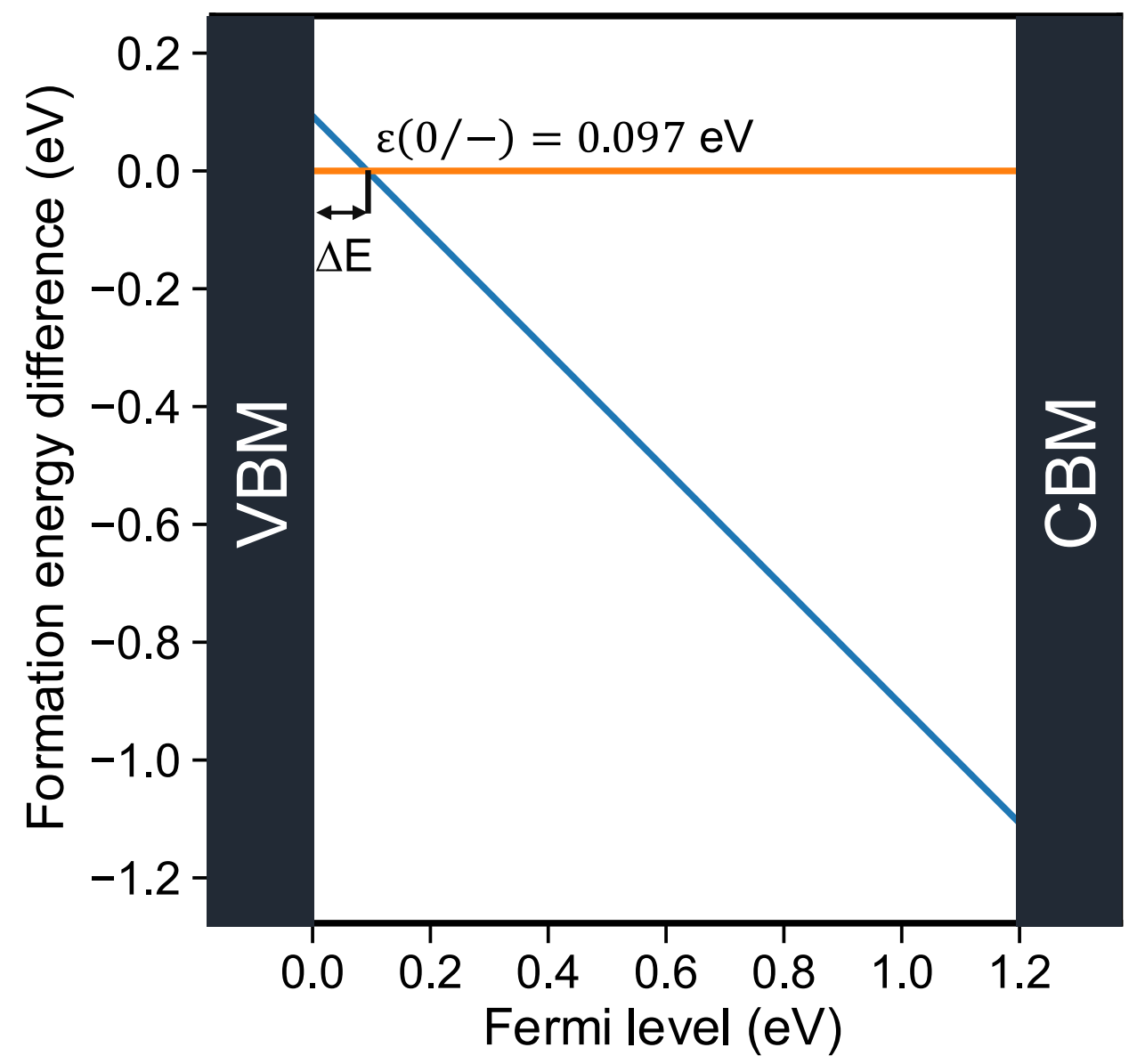

**Figure S3 Formation energy diagram for extracting the charge transition level.** The formation energies between the neutral excited state (orange) and the electron-capture state with one additional electron ($q = -1$, blue) are plotted as a function of the Fermi level. Their crossing point gives the charge transition level, $\varepsilon(0/-)$, which corresponds to $\Delta E = 0.097$ eV and is used in the nonradiative lifetime calculation.

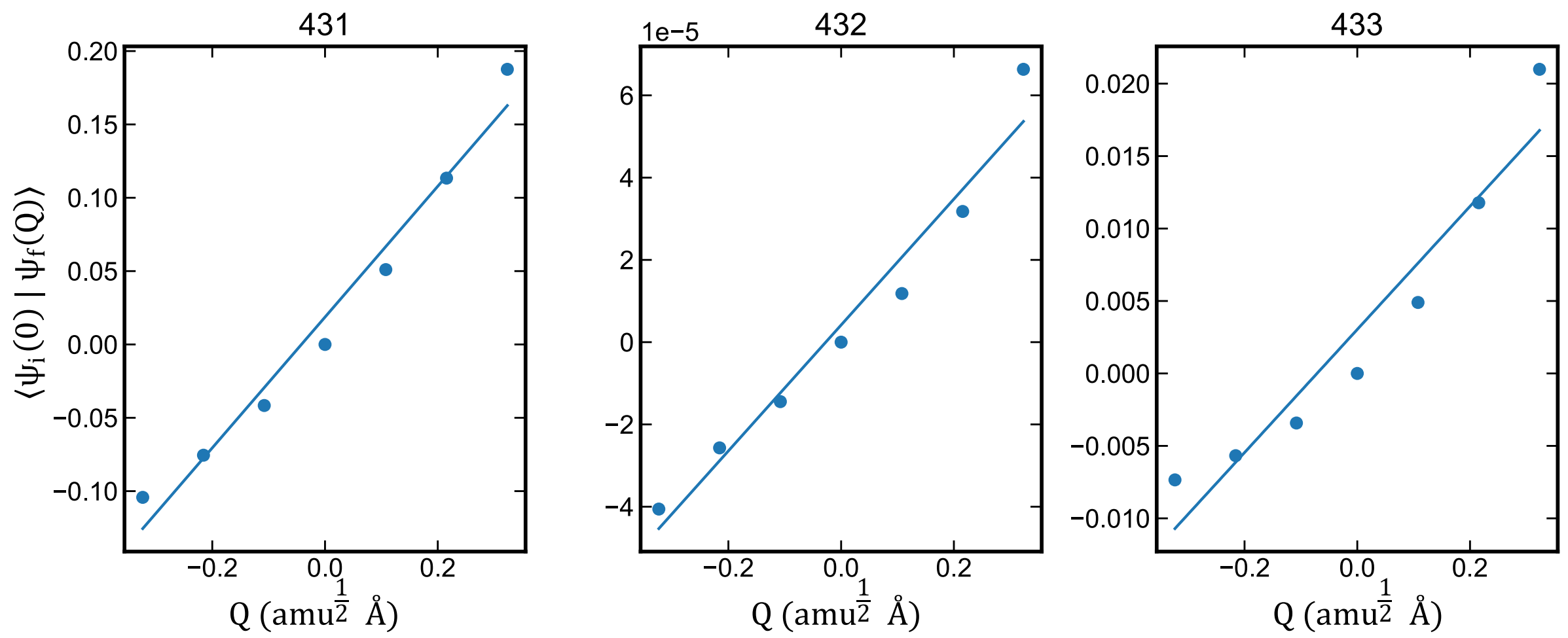

**Figure S4 Calculation of electron-phonon coupling matrix elements.** Wave-function overlaps $\langle \psi_i(0) \mid \psi_f(Q) \rangle$ between the relaxed structures and each displaced structure as a function of the generalized coordinate, $Q$[1,2]. The derivative of this overlap is used to calculate the coupling matrix elements, $W_{\mathrm{if}}$. Since the silicon valence band maximum consists of three nearly degenerate bands (band indices 431–433), the overlaps are computed separately for each band, and their average is used in the calculation of the capture coefficient.

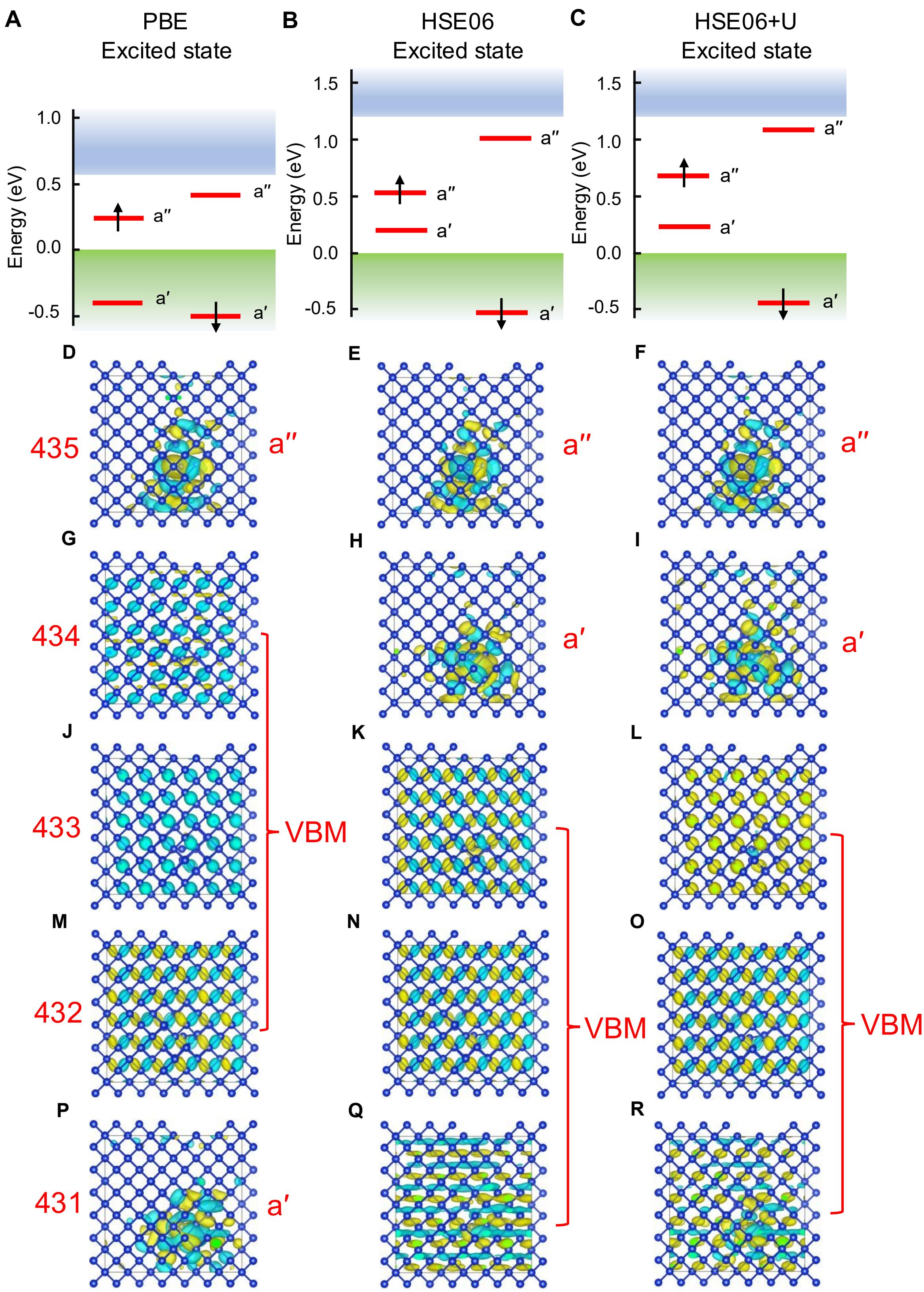

**Figure S5 Wavefunctions of the defect orbitals**. Kohn-Sham energy levels for the excited

state calculated by the PBE functional in **A**, and calculated by HSE06 functional without $U$ in **B** and with $U$ in **C**. Bottom panels show wavefunctions for each band level (only up-spin orbitals). The VBM is a superposition of three nearly degenerate bands (band indices 432–434 in **A**, 431–433 in **B** and **C**). Across all states, the a′ defect levels remain localized and do not hybridize with the continuum bands.

**Table S1. Energy differences by different functionals.** Zero-phonon line (ZPL) energies and the charge transition levels, defined as $\Delta E = E_{\mathrm{cs}} - E_{\mathrm{ex}}$, evaluated using different functionals. The total energies of the ground, excited, and electron-capture states are denoted by $E_{\mathrm{gs}}$, $E_{\mathrm{ex}}$, and $E_{\mathrm{cs}}$, respectively. While PBE and HSE06 yield negative $\Delta E$, HSE06+U results in a positive $\Delta E$, reflecting the exchange-driven energy-level reordering.

| Functionals | ZPL $= E_{\mathrm{ex}} - E_{\mathrm{gs}}$ (eV) | $E_{\mathrm{cs}} - E_{\mathrm{gs}}$ (eV) | $\Delta E = E_{\mathrm{cs}} - E_{\mathrm{ex}}$ (eV) |
|---|---|---|---|
| PBE | 1.149 | 0.430 | -0.719 |
| HSE06 | 0.936 | 0.931 | -0.005 |
| HSE06+U | 0.953 | 1.050 | 0.097 |